\documentstyle[preprint,aps,psfig]{revtex}
\def\BA{\begin{eqnarray}}
\def\BE{\begin{equation}}
\def\EA{\end{eqnarray}}
\def\EE{\end{equation}}
\def\eps{\varepsilon}

\def\gtsim{\lower-0.45ex\hbox{$>$}\kern-0.77em\lower0.55ex\hbox{$\sim$}}
\def\ltsim{\lower-0.45ex\hbox{$<$}\kern-0.77em\lower0.55ex\hbox{$\sim$}}
\draft
\begin{document}
\title{Critical Correlations of Wilson Lines in $SU(3)$ and the
High Energy $\gamma^*p$ Cross Section}
\author{Hans J.~Pirner$^{ab}$
\footnote{pir@tphys.uni-heidelberg.de}
 and Feng Yuan$^a$
\footnote{yf@tphys.uni-heidelberg.de}}
\address{${}^a$Institut f\"ur Theoretische Physik der
Universit\"at Heidelberg, Germany}
\address{${}^b$ Max-Planck-Institut f\"ur Kernphysik Heidelberg, Germany }
\maketitle
\begin{abstract}
We discuss deep inelastic scattering at high energies  as a
critical phenomenon in $2+1$  space - time dimensions.  In the
limit of Bjorken $x \rightarrow 0$, $QCD$ $SU(3)$ with quark
fields becomes a critical theory  with a diverging correlation
length $\xi(x) \propto x^{-\frac{1}{2 \lambda_2}}$ where the
exponent $\lambda_2=2.52$ is obtained from the center group $Z(3)$
of $SU(3)$. We conjecture that the dipole wave function of the
virtual photon for transverse sizes $1/Q<x_{\bot}<\xi$ obeys
correlation scaling $\Psi \propto (x_{\bot})^{-(1+n)}$ before
exponentially decaying for distances larger than the correlation
length. Using this behavior combined with different $x$
-independent dipole proton cross sections we calculate the proton
structure function and compare with the experimental data. We
take the good agreement with the measured proton structure
function F$_2(x,Q^2)$ as an indication that at high energies
dimensional reduction to an effective three dimensional theory
with a critical point occurs. \vspace{1pc}
\end{abstract}

\newpage
\section{Introduction}
Deeply inelastic electron-proton scattering at very high energies
has shown that the cross section of the virtual photon at high
virtuality $Q^2$ increases faster with energy than at low
virtuality. Small size objects experience a stronger energy
dependent cross section than large size objects. Perturbative
$QCD$ has been partially successful in  explaining this physics.
With suitable starting distributions  next to leading order DGLAP
- evolution can reproduce the experimental data. NNLO analysis
\cite {Martin:2002dr} shows a slight improvement. But at small
$x$ the DGLAP- summation of $log(Q^2)$ terms and neglect of
$log(1/x)$ terms is theoretically unsatisfactory. Therefore there
has been a considerable effort \cite
{BFKL,Bal,Alt,ABF,Mueller,Bartels} to include the leading
$log(1/x)$  contributions. It has been suggested \cite {Bal} that
the relevant collective variables at high energies are the gluon
phase factors (Wilson lines) collected by the low $x-$ partons in
the photon on near light like trajectories. We have followed
this  promising method in Ref.~\cite{PirnerPLB} relating high
energy scattering to the behavior of Wilson line correlations in
a $2+1$ dimensional Hamiltonian near the light cone
\cite{NPFV,HJP}. In this Ref. \cite {PirnerPLB} the dynamics of
Wilson lines in $QCD$ with two colors ($N_c=2$) has been
considered. In the present  paper we address the realistic case
of SU($N_c$) with three colors ($N_c=3$). We  propose that with
increasing energy the effective dynamics reaches a critical point
which is characterized by the symmetry properties of $SU(3)$.
Universality tells us that $SU(3)$ Wilson lines have the same
correlation functions as Potts spins in the center group $Z(3)$ in
three dimensions. Since the correlation functions of Wilson lines
influence the structure of the photon wave function at high
energies or small $x$, the $Z(3)$ symmetry determines the strong
increase of the $\gamma^*p$ cross section at high energies.

The discovery of asymptotic freedom \cite {Wilczek,Politzer}
gives a fixed point of $SU(3)$ at $Q^2 \rightarrow \infty$, which
has opened up the possibility of perturbative calculations in
$QCD$. At infinitely high energies the longitudinal momentum
transfers are minute, therefore $QCD$ at high energies reduces to
a $2+1$ dimensional theory, when the longitudinal space variables
have been integrated out. Only transverse dynamics matters. The
resulting effective Hamiltonian has a fixed point at $1/x
\rightarrow \infty$. Lattice simulations for $SU(2)$ \cite {HJP}
have found an effective coupling at this fixed point which is not
small. These lattice simulations still have to be extended to
$SU(3)$. Therefore in the following paper we will use mostly
symmetry arguments to characterize the physics near this critical
point. This critical point influences the physics whenever a
large number of low $x$ partons is involved. Beyond virtual
photon-proton scattering other testing grounds for critical point
dynamics will be RHIC and LHC.

As framework we use the formalism of a near light cone
Hamiltonian. The light cone formulation of $QCD$ has been a useful
tool in perturbative calculations and extended beyond. Since the
vacuum in the strictly light cone formulation is simple, the
Hamiltonian must include all the complicated structures of $QCD$
like condensates. Therefore we prefer a theoretical form
developed in Refs. \cite {NPFV,Prokhvatilov,Lenz2} where the
light cone is reached in a limiting procedure and quantization is
always on spacelike surfaces. We start by choosing the following
$\eta-$ coordinates which smoothly interpolate between the
Lorentz and light front coordinates. The parameter $\eta$
specifies how near to the light cone the coordinate system is: \BA
x^t =x^{+} &=& \frac1{\sqrt2} \left\{ \left(1 + \frac{\eta^2}{2}
\right) x^{0} + \left(1 - \frac{\eta^2}{2} \right) x^{3}
\right\} ,\nonumber \\
x^{-} &=& \frac1{\sqrt2} \left( x^{0}-x^{3} \right)~.
\label{Coor}
\EA
For high energy photon proton scattering at small $x=Q^2/s$,
with $s=W^2$ as  c.m. energy squared  we define two light like vectors
using the photon vector
$q, q^2=-Q^2$ and the proton vector $p,p^2=m^2 \approx 0$ .
\BA
e_1 &=& q- \frac{q^2}{2 pq}p\nonumber \\
e_2 &=& p.
\EA
For finite energies the vector of the photon $q$ which points in the
direction
of $x^t$
can be calculated as a
linear combination of the light like vector $e_1$ with a small amount
of $e_2$ admixed
\BA
e_{\eta} &=& q +xp -\frac{\eta^2}{2} p\nonumber \\
         &=& e_1-\frac{\eta^2}{2}e_2.
\EA
One sees that the mixing $\eta$
is related to the
Bjorken variable $x$ and vanishes in the limit of infinite energies
as $\frac{\eta^2}{2}=x$.
Therefore it is natural to formulate high energy scattering
in near light cone coordinates.
In vacuum it is appropriate to use the Hamiltonian with
periodic boundary conditions.
In this work we only study the effective quark density function
of the photon,
i.e. a vacuum problem without theoretically modeling the proton target.

The outline of the paper is as follows: In Section II we give the near
light cone Hamiltonian and its reduction to $(2+1)$ dimensions for
small $x$. Section III is devoted to the approach of the critical
point. Section IV gives the behavior of the correlations of Wilson lines
near the critical point. In Section V this correlation length is
used to model the effective quark density in the photon and calculate
the structure function $F_2$ of the proton. A comparison to the
HERA-data is given in the same section.
Section VI contains the  conclusions.

\section{Near light cone $SU(3)$ $QCD$ Hamiltonian}

For small $x$ the eikonal phases acquired by the
quarks/antiquarks are the relevant collective variables. The
light cone Hamiltonian on the finite light like $x^-$ interval of
length $L$ has Wilson line or Polyakov operators similarly to
$QCD$ formulated on a finite interval in imaginary time at finite
temperature \BE P(\vec{x}_\bot) = \frac1{N_c} \mbox{tr~P exp}
\left (ig \int\limits_0^L    dx^-A_-(\vec{x}_\bot,x^-)\right ).
\label{Polyakov} \EE

The dynamics of these Polyakov operators is determined by the
near light cone Hamiltonian $H$ which has been derived in
reference \cite {NPFV}. In the Weyl gauge $A_+=0$, the Gauss-law
constraint can be resolved for $\Pi_-$ and one obtains a
Hamiltonian which depends   on the zero mode momentum $p_-$. The
subsequent  modified light front gauge $\partial_-A_-=0$
eliminates $x^-$ dependent fields $A_-(x^-,x_{\bot})$, but allows
fields $a_-(x_{\bot})$ which are functions of the transverse
coordinates only. The final $QCD$ Hamiltonian near the light cone
has the following form \cite {NPFV}:
\begin{equation}
H  = \int dx^- dx_{\bot} {\cal H}(x_{\bot}, x^-) \ ,
\label{Ham}
\end{equation}
with
\begin{eqnarray}
{\cal H} & = & \mbox{tr} \left[  \partial_1 A_2 -\partial_2 A_1
-ig [A_1, A_2]  \right]^2
 +  \frac{1}{\eta^2} \mbox{tr} \left[
\Pi_{\bot}-\left(\partial_-A_{\bot}-
ig[a_-,A_{\bot}]\right)\right]^2 \nonumber \\
& + &  \frac{1}{\eta^2} \mbox{tr} \left[ \frac{1}{L} e_{\bot}
-
\nabla_{\bot} a_-\right]^2
 +  \frac{1}{2 L^{2}}  p_{-}^
{\, \dagger}(x_{\bot})
p_{-}(x_{\bot})  \nonumber \\
& + &   \frac{1}{L^{2}}\int_{0}^{L} dz^{-} \int_{0}^{L} dy^{-}
\sum_{p,q,n}\,^{'}\frac{ G_{\bot qp}(x_{\bot}, z^{-})
G_{\bot pq}(x_{\bot},y^{-})}{\left[ \frac{2\pi n}{L} +
g(a_{-q}(x_{\bot})- a_{-p}(x_{\bot})) \right]^{2}}
e^{i2\pi n(z^{-}-y^{-})/L} \nonumber\\
& - & i\frac{2}{\eta^2}\psi^{\dagger}_-(\partial_- -ig a_-)\psi_- -
i\frac{1}{\eta}\psi^{\dagger}(\alpha_{\bot}-ig A_{\bot})\psi
+m\frac{1}{\eta}\psi^{\dagger}\beta\psi ,
\label{Hamil}
\end{eqnarray}
The term $e_{\bot}$ depends on an external source.
\BE
e_{\bot}^{c_0}=g \nabla_{\bot}\int dy^- dy_{\bot}\\
\frac{1}{\nabla_{\bot}^2}
\left(f^{c_0 a b }A_{\bot}^a(y_{\bot},y^-) \Pi_{\bot}^b(y_{\bot},y^-)\\
+\rho^{c_0}(y_{\bot},y^-)\lambda^{c_0}/2 \right ). \EE In the $x
\rightarrow 0$ limit, those pieces ${\cal H^{\eta} }$ of the
Hamiltonian ${\cal H }$ dominate which are most singular at
$\eta=0 $ and do not couple to the three dimensional gauge fields
$A_{\bot}$ and $\psi_+$. This reduced Hamiltonian has collective
variables $a_-^{c_0}$ with the color indices $c_0=3,8$ \BE
a_-^{c_0}=\frac {1}{L} \int\limits_0^L
dx^-A_-^{c_0}(\vec{x}_\bot,x^-) \EE which determine the Wilson
lines and live in a $2+1$ dimensional space. The Wilson line
operators $ P(x_{\bot})$ can always be parameterized in terms of
the diagonal color matrices $a_-^3\lambda_3/2$ and
$a_-^8\lambda_8/2$   by suitable gauge transformations. We
consider the dynamics of the fields $a_-$ in vacuum, i.e. without
the source term $e_{\bot}$. A possible external source, e.g., a
Gaussian distributed random color charge \cite {I}, can be taken
into account via this term which shifts the zero mode fields
$\nabla_{\bot} a_-^{c_0}$ to fluctuate around the classical
fields $e_{\bot}^{c_0}$. We recall that the $\eta-$ coordinates
correspond to the physics in a fast moving frame and factorize
the reduced energy from the Lorentz boost factor $\propto
\frac{1}{\eta}$ and a  transverse lattice cut off $a$ \BA
\nonumber
h_{red} &=& 2 \eta a \int dx^- dx_{\bot} {\cal H^{\eta}}\\
        &=&\int  dx^- dx_{\bot} \sum_{c^0=3,8}
\left( \frac{2 a}{\eta} tr(\frac{1}{L}e_{\bot}^{c_0}-
\nabla_{\bot}a_-^{c_0})^2+\frac{2 a \eta}{2 L^2}p_-^{c_0 \dagger}
p_-^{c_0} -\frac{4 a}{\eta}\psi^{\dagger}_-~g a_-^{c_0}
\frac{\lambda^{c_0}}{2} \psi_-\right).
\EA
Then we redefine modified zero mode fields and their conjugate momenta
on a transverse lattice at positions $b_{\bot}$:
\BA
\nonumber
\varphi^{c_0} (\vec b_\bot)&=&\frac{1}{2} g L a_-^{c_0}(b_{\bot}),\\
\frac{\delta}{\delta \varphi^{c_0}(b_{\bot})}&=&a^2
\frac{\delta}{\delta \varphi^{c_0}(x_{\bot})}. \EA Their dynamics
on the lattice is determined by the lattice Hamiltonian $h_{lat}$
(cf. also Ref. \cite{HJP} for the case of $SU(2)$), \BA \nonumber
  h_{\rm lat} &=&   \sum_{\vec b,c_0}[-g^2_{\rm eff}
    \frac1J \frac\delta{\delta\varphi^{c_0}(\vec b)}
          J \frac\delta{\delta\varphi^{c_0}(\vec b)}\\
   &&+ \frac{1}{g^2_{\rm eff}} \sum_{\vec\eps}
    \left( (\varphi^{c_0}(\vec{b})-\varphi^{c_0}(\vec{b}+\vec\eps))^2
\right ) -\frac{4 a}{\eta L}<\psi^{\dagger}_-\psi_->^{c_0}
\varphi(\vec{b})^{c_0}]. \label{hhat}
\EA
The summation goes over
color indices $c_0=3,8$ and all $2$ -dimensional lattice sites.
The geometry of the $SU(3)$ manifold enters via the Jacobian \cite
{Lenz1,beg}: \BE J=\sin ^2(\varphi^3)\sin
^2(\frac{1}{2}(\varphi^3-\sqrt{3}\varphi^8)) \sin
^2(\frac{1}{2}(\varphi^3+\sqrt{3}\varphi^8)). \EE Because of
dimensional reduction the effective coupling constant depends on
the length $L$ of the interval in $x_-$ direction and the
$QCD$-coupling $g^2$ multiplied by the parameter $\eta$
characterizing the nearness to the lightcone 
\BE 
\label{geff}
 g^2_{\rm eff} = \frac{g^2L\eta}{4a}.
\EE
The phase factors $\varphi^3,\varphi^8$ are defined in the
fundamental domain  (cf. Fig.~1)
$$
\begin{array}{rcccl}
  0                  & \leq & \varphi^3 & \leq & \pi \\
  -\varphi^3/\sqrt{3}& \leq & \varphi^8 & \leq & \varphi^3/\sqrt{3}.
\end{array}
$$
The Jacobian vanishes on the boundaries of the fundamental domain.
The coupling to the quarks is introduced by the external field \BE
<\psi^{\dagger}_-\psi_->^{c_0}=\int dx^- \psi_-^{\dagger c_0}
\psi_-^{c_0}. \EE It was shown in perturbation theory \cite
{Burkardt1} that the product of ``bad'' light cone components
$\psi_-$ has  a zero mode contribution. In structure functions
such a singular piece is generated at Bjorken $x=0$. In $QCD$ on
the light cone \cite {Brodsky,Burkardt2} the negative energy
states obey a constraint equation which allows a constant $x^-$
independent solution. Therefore we assume that the fermion
negative energy states develop an expectation value
$<\psi^{\dagger}_-\psi_->^{c_0}$. Further work is still needed to
show how such a nonvanishing zero mode density
$<\psi^{\dagger}_-\psi_->^{c_0}$ is realized near the light cone.

In the continuum limit of the transverse lattice theory the
lattice size $a$ goes to zero and/or the extension $L$ of the
lattice to infinity. This limit combined with the light cone
limit $\eta \rightarrow 0$ leads to an indefinite behavior of the
effective coupling constant (cf.~Eq.~(\ref{geff})). The critical
behavior of the zero-mode theory resolves this ambiguity. In Ref.
\cite {HJP} we have done a Finite Size Scaling (FSS) analysis for
$SU(2)$ $QCD$ obtaining a second order transition as a function
of the coupling $g^2_{\rm eff}$ between a phase with massive
excitations at strong coupling and a phase with massless
excitations at weak coupling. In the strong coupling domain of
$g^2_{\rm eff}$ the energy of the rotators $\varphi^{c_0}$ is
dominated by the electric energy $\propto g^2_{eff} \frac1J
\frac\delta{\delta\varphi^{c_0}(\vec b)} J
\frac\delta{\delta\varphi^{c_0}(\vec b)}$ which corresponds to the
Laplacian in the group manifold. Each site has an energy spectrum
with a gap $\eps_n=n(n+2)\eps_0$ in $SU(2)$ or
$\eps_n=n(n+4)\eps_0$ in $SU(3)$ \cite {beg}. With decreasing
$g^2_{\rm eff}$ the magnetic coupling $\propto \frac{1}{g^2_{\rm
eff}}(\varphi^{c_0}(\vec{b}) -\varphi^{c_0}(\vec{b}+\vec\eps))^2$
becomes stronger. A larger  nearest neighbor coupling leads to a
coherently aligned  ground state which has massless excitations.
Consequently the mass gap vanishes at a sufficiently small
$g^2_{eff}$. The resulting critical $SU(2)$ theory is in the same
universality class as the $Z(2)$ theory or the Ising model in
$3$-dimensions, which has been checked  in the lattice simulations
\cite {HJP} with the available numerical accuracy. Barring any
unusual effects arising from the coupling to the three
dimensional gluon fields, we think that the full theory may shift
the exact value of the critical coupling, but does not change the
critical exponents which are determined by the symmetry of the
problem. The Abelian collective fields in $SU(2)$ have reflection
symmetry around $\pi/2$ which corresponds to the up-down symmetry
of the Ising spins.

\section{How to reach the critical point in $SU(3)$ at small $x$ }

The reduced $SU(3)$ Hamiltonian has rather  different symmetry
properties than the $SU(2)$ Hamiltonian.  A closer look at the
fundamental domain in $SU(3)$ (cf.~Fig.~1) shows three regions
separated by dashed lines. An element from the center group $Z(3)$
can be mapped from one region to another by large gauge
transformations. In $SU(2)$ these large gauge transformations are
reflections around $\pi/2$ on the $\varphi^3$ axis. In previous
work \cite {HJP} we found that the universality class of the zero
mode $SU(2)$ theory is the same as $Z(2)$ theory, therefore we
think that the universality class of the reduced Hamiltonian in
$SU(3)$ is the three state Potts model $Z(3)$.  In each subregion
of the fundamental domain the zero mode variables
$\varphi^3,\varphi^8$ are represented by one spin orientation.
The relevant center group $Z(3)$ has a  weak first order
transition whose critical line ends in a second order point in
the presence of an external field. We conjecture that this
external field is provided by the fermion zero mode density near
the light cone.

To match the Hamiltonian lattice  with scattering we consider the
same physical picture as in  Ref. \cite {PirnerPLB}. The lattice
constant $a$ is chosen to coincide with the photon resolution $
\approx \frac{1}{Q}$. The longitudinal extension $L$ must be
larger than the color coherence length of the $q \bar q $ state
in the photon-proton c.m. system. The  photon and proton move on
almost light like trajectories and the coherence length grows
with $x$ as: \BE \Delta x^-= \frac {1}{Q \sqrt {x}}\ . \EE We
demand therefore that $\frac{L}{a} >  \frac {1}{Q a \sqrt{x}}$.
In the limit of small $x$ the parameter $\eta$ decreases as
$\eta=\sqrt{2x}$. Assuming that a stable fixed point in the
effective   running coupling $g_{eff}^2=g^2\frac{L \eta}{4 a}$
exists, we get \BE \frac{L}{a} \approx\frac{c}{g^2 \sqrt{x}}. \EE
Since $1/a \approx Q$ , we can choose a positive constant $N_0$
such that the interval $L$ exceeds the color coherence length of
the $q \bar q$ state in the photon, namely 
\BE
\label{e17}
\frac{L}{a} = \frac{1}{g^2}(N_0+\frac{c}{a Q
\sqrt{x}}). 
\EE 
In $Z(3)$ theory the critical behavior depends on
two parameters, the effective field $\hat h$ and the coupling
$\hat \tau$: Using equation (\ref{e17}) one finds
that the external field $h= \frac{4 a
<\psi_-^{\dagger}\psi_->}{\eta L}$ converges in the limit $x
\rightarrow 0$ to $h^*$: \BE h^*= \frac{4 g^2
<\psi_-^{\dagger}\psi_->}{\sqrt{2} c}. \EE The limiting external
field $h^*$ is independent of the short distance cutoff $a$ under
the assumption that the light cone density of the quarks
$<\psi_-^{\dagger}\psi_->$ behaves in the same way with $a$ as the
perturbative quark density in the photon namely $\propto \int d
r_{\bot}^2 K_1^2( r_{\bot}^2)$. It is specific to high energy
scattering that the external field and the coupling are
approaching their critical values in a similar way for $x
\rightarrow 0$ \BA
\hat h=|h-h^*| \approx h_0 \sqrt{x},\\
\hat \tau= \frac{g^2 L\eta }{4 a}- g^{* 2}_{eff}
\approx  t_0 \sqrt{x}.
\EA
It would be important  to
find another physical situation where the experimental conditions
regulate both quantities
independently.

\section{Behavior of the correlation length near the critical point}

The earliest work on critical behavior of pure $SU(3)$ gauge
theory \cite {Svetitsky} has been done in the context of the
finite temperature phase transition. Recent studies \cite
{Karsch1,Karsch2} have established  common features of  $SU(3)$
$QCD$ and the three state Potts model $Z(3)$ with a first order
phase transition line which ends in a critical point where the
transition is second order.  In Refs. \cite {Karsch1,Karsch2} the
universality class has been identified as the 3-dimensional Ising
model in three dimensions which has the critical exponents :
\begin{eqnarray}
\nonumber
 \lambda_1 &=& \nu^{-1}      = 1.56,\\
 \lambda_2 &=& d -\beta \lambda_1 = 2.52.
\nonumber
\end{eqnarray}

In order to assess the divergence of the
correlation length,
we rescale the lattice constant by a factor
$b$ and the external field and coupling with powers of $b$
according to their
respective anomalous dimensions
\BE
\frac{\xi}{a} = b f(b^{\lambda_1}\hat \tau,b^{\lambda_2} \hat h) .
\EE
When the external field $\hat h \rightarrow 0$
and $\hat \tau \hat h^{-\lambda_1/\lambda_2}$ is sufficiently
close to zero, the critical behavior of the
correlation length is calculable by
choosing $b^{\lambda_2} \hat h=1$, then one obtains
\BA
\nonumber
\frac{\xi}{a} &=& \hat h^{-1/\lambda_2}
f(b^{\lambda_1}\hat \tau,1)\\
    &=& \hat h^{-1/{\lambda_2}} f_h(\hat \tau
\hat h^{-\lambda_1/{\lambda_2}}). \label{e22}
\EA
The function
$f_h(r)=f(r,1)$. Note the power ${1}/{\lambda_2}=0.4$ governing
the low $x$ behavior of the correlation length is considerably
smaller than the power ${1}/{\lambda_1}=0.63$ which determines
the power for a small ratio $\hat h \hat
\tau^{-\lambda_2/\lambda_1}$. In the latter case one gets with
the function $f_{\tau}(s)=f(1,s)$ the behavior \BA \nonumber
\frac{\xi}{a} &=& \hat \tau^{-{1}/{\lambda_1}} f(1,\hat h \hat
\tau^{-{\lambda_2}/
{\lambda_1}})\\
              &=&\hat \tau^{-{1}/{\lambda_1}}f_{\tau}
(\hat h \hat \tau^{-{\lambda_2}/ {\lambda_1}}). \EA

In the previous work on $SU(2)$ we considered a second order
transition in the absence of an external field. Since the
universality class was also of the Ising type a critical behavior
of the correlation length $\xi \propto \tau^{-{1}/{\lambda_1}}$
resulted. Such a power gave a qualitative description of the
scarce data on the longitudinal structure function \cite
{PirnerPLB}, but definitely cannot describe the F$_2$- data as
shown in Ref.\cite {pirner2}.

In the mathematical framework of the $SU(3)$ critical theory
exposed here, it is clear that the growth of the correlation
function with $\hat h^{-{1}/{\lambda_2}} $ is realized, since for
small $x$ the argument $r$  of the scaling function $f_h$ is
small: $r=\hat \tau \hat h^{-{\lambda_1}/{\lambda_2}}\propto
x^{0.18}$. The argument of $f_{\tau}$, however, $ \hat h \hat
\tau^{-{\lambda_2}/{\lambda_1}} \propto x^{-0.3}$ is large for
small $x$. Therefore we can well approximate the argument of
$f_h$ by $r=0$ for high energy scattering. Recently the scaling
function $f_h$  has been calculated in lattice simulations \cite
{engel} and found to become constant at $r=0$. Substituting $\hat
h \propto x^{1/2}$ in Eq.~(\ref{e22}) one finds that the
correlation length $\xi$ increases with $x \rightarrow 0$ as \BE
\xi \propto (\frac{x}{x_0})^{-\frac{1}{2 \lambda_2}} f_h(0). \EE

Near the critical point the Wilson lines experience long range
correlations which means that dipoles in the photon wave function
are correlated over large distances. For finite correlation
length there exists an intermediate range where
${1}/{Q}<x_\bot<\xi$ for which the correlation function of Wilson
lines is power behaved: \BE \label{e17p} <P(x_\bot) P(0)> \approx
\frac{1}{x_\bot^{1+n}} \EE where $n=0.04$ in Ising like systems.
This region is responsible for the well known effect of critical
opalescence in the gas liquid transition. For larger distances
$x_\bot>\xi$ the correlation function decreases exponentially \BE
\label{e18} <P(x_\bot) P(0)> \approx e ^{-{x_\bot}/{\xi}}. \EE

\section { Proton Structure Function $F_2$ and Critical Behavior}

The photon is ideal to investigate high energy cross sections of
hadronic objects with a variable size. In the course of
$x$-evolution the photon wave function develops many dipoles
which in general diffuse into distance scales beyond the original
size $1/Q$. This increase in parton density and/or  size of the
photon wave function is generally believed to be the origin of
the increasing high energy cross section. In this work, we do not
follow the development of the photon dipole state in detail, we
only give a qualitative description of the effective photon size
as a function of $x$ using the results of the $2+1$ dimensional
critical $QCD$ $SU(3)$ theory as a guiding principle. We
parameterize the longitudinal and transverse photon probability
densities as: \BA \nonumber \rho_{\gamma}^T&=&\frac{6 \alpha}{4
\pi^2}\sum_f  \hat e_f^2 \varepsilon^2 [z^2+
(1-z)^2]F_T(\eps x_\bot),\\
\rho_{\gamma}^L&=&\frac{6 \alpha}{4 \pi^2}\sum_f  \hat e_f^2 4 Q^2 z^2
(1-z)^2 F_L(\varepsilon x_\bot),\\
\varepsilon&=&\sqrt{Q^2 z(1-z)}. \EA The perturbative scale for
the photon quark density is given by ${1}/{\varepsilon}$. We
modify  the photon wave function depending on the relation of the
correlation length $\xi$ of the Wilson loops to the perturbative
scale ${1}/{\varepsilon}$. We set \BE \xi= \frac{1}{\varepsilon}
(\frac {x}{x_0})^{-\frac{1}{2 \lambda_2}}\ . \EE For the
reference Bjorken parameter $x_0=10^{-2}$ the correlation length
is fixed at the perturbative scale. The critical exponent
$\frac{1}{2 \lambda_2}=0.2$ determines the Wilson line
correlations for $x<x_0$. If the transverse size of the dipole is
smaller than the perturbative length scale
$x_{\bot}<{1}/{\varepsilon}$ we use  the  perturbative quark
densities $F_{T}(\varepsilon x_\bot)= K_{1}(\varepsilon x_\bot)^2
$ and $F_L(\varepsilon x_\bot)= K_0(\varepsilon x_\bot)^2 $. This
is the uninteresting case. For ${1}/{\varepsilon}<x_\bot< \xi$ we
modify  the perturbative photon density using the correlation
functions of the critical theory, Eqs.~(\ref{e17p},\ref{e18}), \BA
\nonumber F_{T/L}(\eps x_\bot) &=& K_{1/0}(\eps x_\bot)^2
~~~~~~~~~~~~
\textrm{ for  } x_\bot <\frac{1}{\varepsilon}, \\
\nonumber
               &=& K_{1/0}(1)^2 (\frac{1}{\eps x_{\bot}})^{2+2 n} ~~~
\textrm{for  }  \frac{1}{\eps} < x_\bot < \xi ,\\
               &=& K_{1/0}(x/\xi)^2 (\frac{1}{\xi \eps})^{2+2 n}~~
\textrm{for  }  x_\bot > \xi. \label{eftl} \EA The above
parameterizations extends the photon density into the scaling
region using the scaling index $n=0.04$. For distances beyond the
scaling region the density decays exponentially. The connections
are made in such a way that the density is continuous at the
respective boundaries of each region. The photon density combined
with an energy independent dipole-proton cross section determines
the structure function $F_2$ and the photon- proton cross section:
\BA F_2(x,Q^2) &=& \frac{Q^2}{4 \pi^2 \alpha }(\sigma_{\gamma
p}^{T,tot}+
   \sigma_{\gamma p}^{L,tot}),\\
\sigma_{\gamma p}^{T/L,tot}&=&\int d^2 x_\bot \int _0 ^1 dz
\rho_{\gamma}^{T/L}(x_\bot,z) \sigma_{dip}(x_\bot)
\EA
First, we use the Golec-Biernat-W\"usthoff \cite {Golec1}
dipole-proton cross section at fixed $x_0=10^{-2}$ with
$\sigma_0=23 mb$ to calculate the proton structure function,
\BA
\sigma_{GBW}(x_\bot,R_0) = \sigma_0 (1-e^{-\frac{x_\bot^2}{4 R_0^2}}),\\
        R_0   = \frac{1}{1 GeV}
(\frac{x_0}{3*10^{-4}})^{0.145}\ . \EA The numerical values
entering the above formulas are taken over from the original
reference \cite {Golec1}. Note,  the value $R_0=0.33 fm$ at
$x_0=10^{-2}$ is independent of $x$. In Fig.~2 we present the
proton structure function $F_2$ measured at HERA \cite {H1,ZEUS}
as a function of $x$ for various $Q^2$. The solid theoretical
curve describes the data rather accurately. One must be cautious
not to overestimate the agreement, since due to the fixed beam
energy the values of $Q^2$ and $x$ are correlated in
electron-proton  scattering. For intermediate values $5 GeV^2
<Q^2<15 GeV^2$ the theory slightly overestimates the data.

If one approximates  the GBW-dipole proton cross section by a
simple quadratic function at small distances $r<2 R_0$ and a
constant function for $r> 2 R_0$ \cite {Golec2}, 
\BE 
\label{sgbw}
\sigma(r) \approx \sigma_0 \left (\frac {r^2}{4 R_0^2} \Theta(2
R_0- r) + \Theta(r-2R_0) \right), 
\EE 
one can estimate  the
photon proton cross section rather simply. Small errors are
introduced if one further neglects the exponentially suppressed
part of the photon density in the integral over large transverse
distances and sets the anomalous dimension $n \rightarrow 0$.
Then  one  can integrate  the dominant transverse cross section
in $F_2$ up to the correlation length $\xi$ analytically. \BE
\sigma_T=\frac{3 \alpha}{\pi} \sum e_f^2 Q^2 \sigma_0\\
\int_0^1 dz(z^2+(1-z)^2)\int_0^{\infty} r dr \frac{z(1-z)}{r^2 \varepsilon^2}\\
\Theta(1-\frac{r^2}{\xi^2}) \frac{\sigma(r)}{\sigma_0}
\EE

Redefining the integration variable as
$r'=r (x/x_0)^{\frac{1}{2 \lambda_2}} $ one obtains the $\gamma^* -p$
cross section as if the original GBW cross section $\sigma(r,R(x))$ with
running $R(x)$ would have been used.
\BE
\sigma_T=\frac{3 \alpha}{\pi} \sum e_f^2 Q^2 \sigma_0\\
\int_0^1 dz(z^2+(1-z)^2)\int_0^{\infty} r' dr'\\
\frac{z(1-z)}{r'^2 \varepsilon^2}
\Theta(1-\varepsilon^2r'^2) \frac{\sigma^{GBW}(r',R(x))}{\sigma_0}
\EE
where
\BA
R(x)&=&R_0(\frac{x}{x_0})^{\frac{1}{2 \lambda_2}},\\
\frac{1}{2 \lambda_2} &=& 0.2. \EA Therefore the agreement with
the data can be understood analytically. The critical theory
gives the phenomenologically obtained power dependence of $R$
with $x$. The favored $x$ -dependence of GBW is in the range
$0.145$ and $0.20$. Without a model for the proton source, it is
not possible to obtain the absolute length $R_0$. The structure
function \cite {Golec3} scales as a function of $Q^2 R_0^2
(x/x_0)^{{1}/{\lambda_2}}$ as can be easily derived for the
simplified dipole cross section given in Eq.~(\ref{sgbw}).

In order to test the sensitivity of our model to the assumed form of the
dipole cross section we also tried the rather different dipole cross section
of Forshaw et al. \cite {Forshaw} with $m_q=0$ which parameterizes the
weak energy dependence of the cross section for large dipoles separately from
the strong energy dependence of the cross section for small dipoles.
\BA
\sigma(s,r)&=&\sigma_{soft}(s,r)+\sigma_{hard}(s,r),\\
\sigma_{soft}(s,r)&=&g_1(r)s^{0.06},\\
\sigma_{hard}(s,r)&=& g_2(r)s^{0.38}. \EA In the expression for
$s=Q^2/x$ we fix $x_0=10^{-2}$. The $Q^2$ dependence in the cross
section remains. These cross sections for different values of
$Q^2$ look rather different from the GBW cross section and its
simplified form (cf.~Fig.~3). Using the same modified photon wave
function inside the scaling region we find the theoretical
(dashed) curves in Fig.~2. In spite of the rather different shape
of the dipole proton cross section the agreement of the theory
with the data is again rather good. In our opinion this
represents a strong indication that the theory does not rely on a
specific choice of the dipole proton cross section. Any
reasonable choice of dipole-proton cross section at fixed $x_0$
should do.

\section{Conclusions}

The presented calculation of the proton structure function
suggests the discovery of a new critical point in $QCD$. Besides
the fixed point of asymptotically vanishing coupling at infinite
momentum transfer $Q^2 \rightarrow \infty$ we propose that $QCD$
has another critical point at infinite energies ${1}/{x}
\rightarrow \infty$ . It is characterized by the exponents of the
center group $Z(3)$ of $SU(3)$ and determines the correlation
functions of Wilson lines near the light cone. A correlation
length which increases as $\xi \propto
(\frac{x}{x_0})^{-\frac{1}{2 \lambda_2}}$ with the critical
exponent $\lambda_2=2.52$ given by the 3-state Potts model $Z(3)$
in three dimensions gives the correct effective photon wave
function to explain the growing photon proton cross sections
observed in HERA \cite {H1,ZEUS}.

This result is independent of the specific form of the dipole
proton cross section which is used as an input at the scale
$x_0=10^{-2}$. For the Golec-Biernat-Wuesthoff \cite {Golec1}
and the Forshaw et al. \cite {Forshaw} cross sections
good agreement can be achieved using the effective photon
dipole density which has been modified according to the
critical scaling Greens function.
In this paper we have made a further step extending the
picture of a dilute parton gas into a liquid like phase
at small $x$. At small $x$
a critically diverging correlation length in transverse space
leads to increasing cross sections in qualitative
agreement with the observed growth of structure functions
for high  $Q^2$ deep inelastic scattering.
In various studies \cite{Lipatov:1990zb,Ewerz:2001uq}
of high energy scattering the
conformal invariance of the resulting effective theory has been
pointed out. In $2+1$ dimensions near the critical point
the effective theory is also conformal invariant,
whether the Greens functions in transverse space show
this symmetry is not clear.

Clearly on the theoretical side it is necessary to take into
account the target at $x^-=0$ in the calculation. As a  next step
in a lattice calculation, we plan to include an external source
representing the proton target and calculate the Wilson line
correlation functions with twisted boundary conditions. Averaging
over a stochastic source distribution is also possible \cite {I}.
After such a calculation one does not have to rely anymore on
parameterized dipole proton cross sections which make the
extraction of the critical dynamics difficult. The agreement so
far between the reduced $QCD$ Hamiltonian near the light cone with
experiment gives a strong motivation to pursue the investigation
of the full Hamiltonian in $SU(3)$.

{\bf Acknowledgments}: We thank our colleagues F. Wegner, R.
K\"uhn, C. Ewerz, M. Thies and F. Lenz for helpful discussion on
the critical system and its possible realizations in the high
energy world. We are grateful to F. Karsch  and especially J.
Engels to communicate results on the scaling function in $Z(3)$
prior to publication.

This work has been partially funded through the European TMR
Contract $HPRN-CT-2000-00130$: Electron Scattering off Confined
Partons and supported by INTAS contract ``Nonperturbative QCD''.


\begin{figure}[t]
\vskip 1cm
\centerline{\psfig{figure=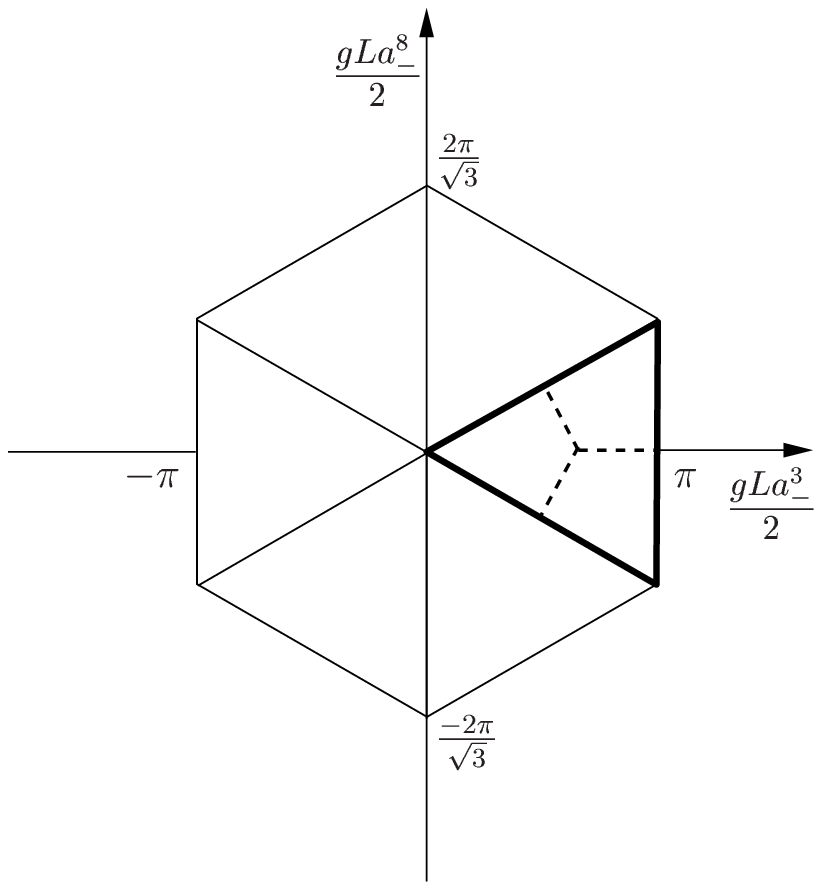,width=8cm,height=8cm}}
{\caption{The fundamental domain of gauge field variables $\varphi_3=
\frac{g L a^3_-}{2}$ and
$\varphi_8=\frac{g L a^8_-}{2}$.}}
\end{figure}

\begin{figure}[t]
\centerline{\psfig{figure=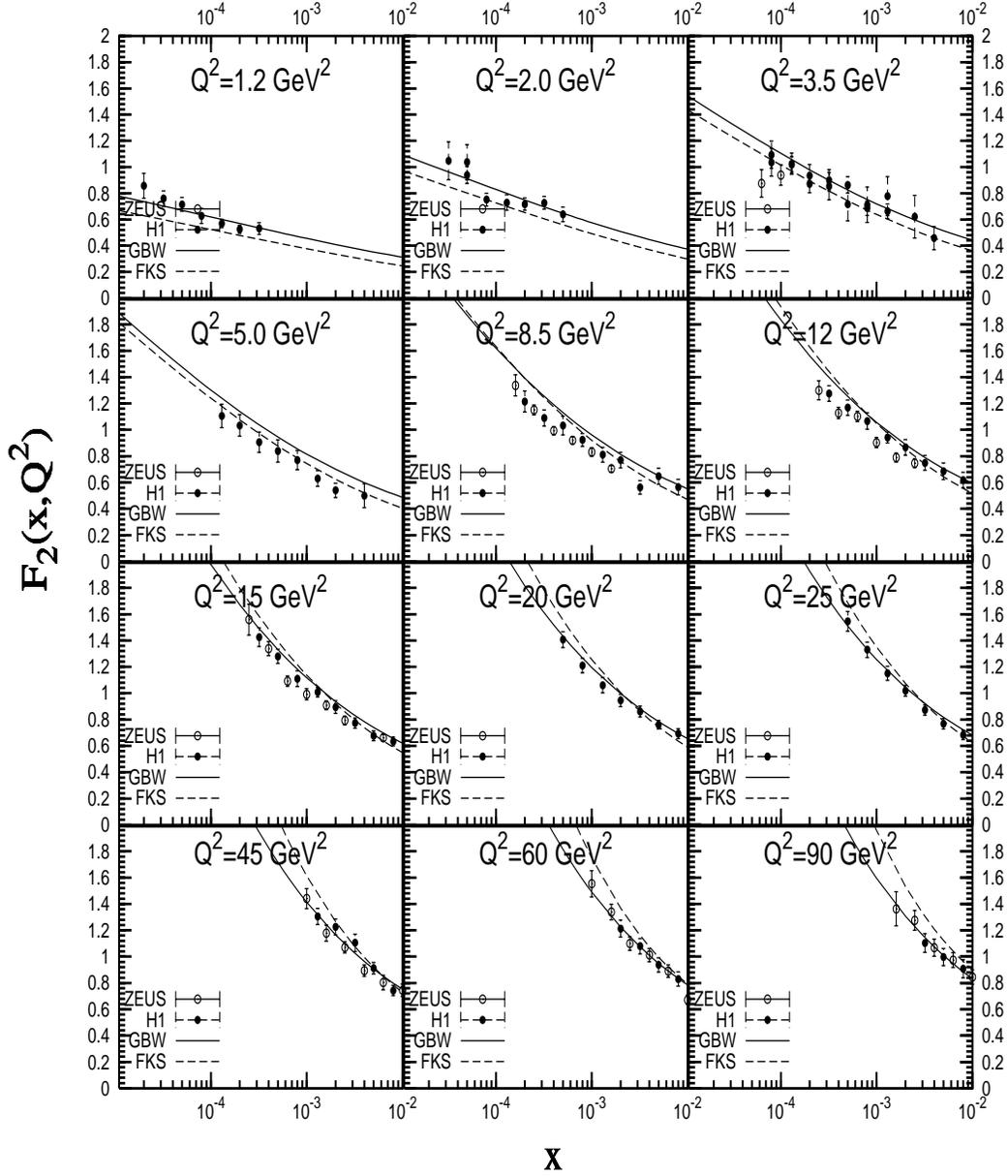,width=14cm,height=18cm}}
{\caption{ Proton structure function $F_2$ as a function of $x$
for various $Q^2$. The experimental data are from H1 and ZEUS at
HERA \protect\cite{H1,ZEUS}. The curves are the theoretical results of
the critical theory with the functions $F_{T/L}(\epsilon x_\bot)$
of Eq.~(\ref{eftl}). }}
\end{figure}

\begin{figure}[t]
\centerline{\psfig{figure=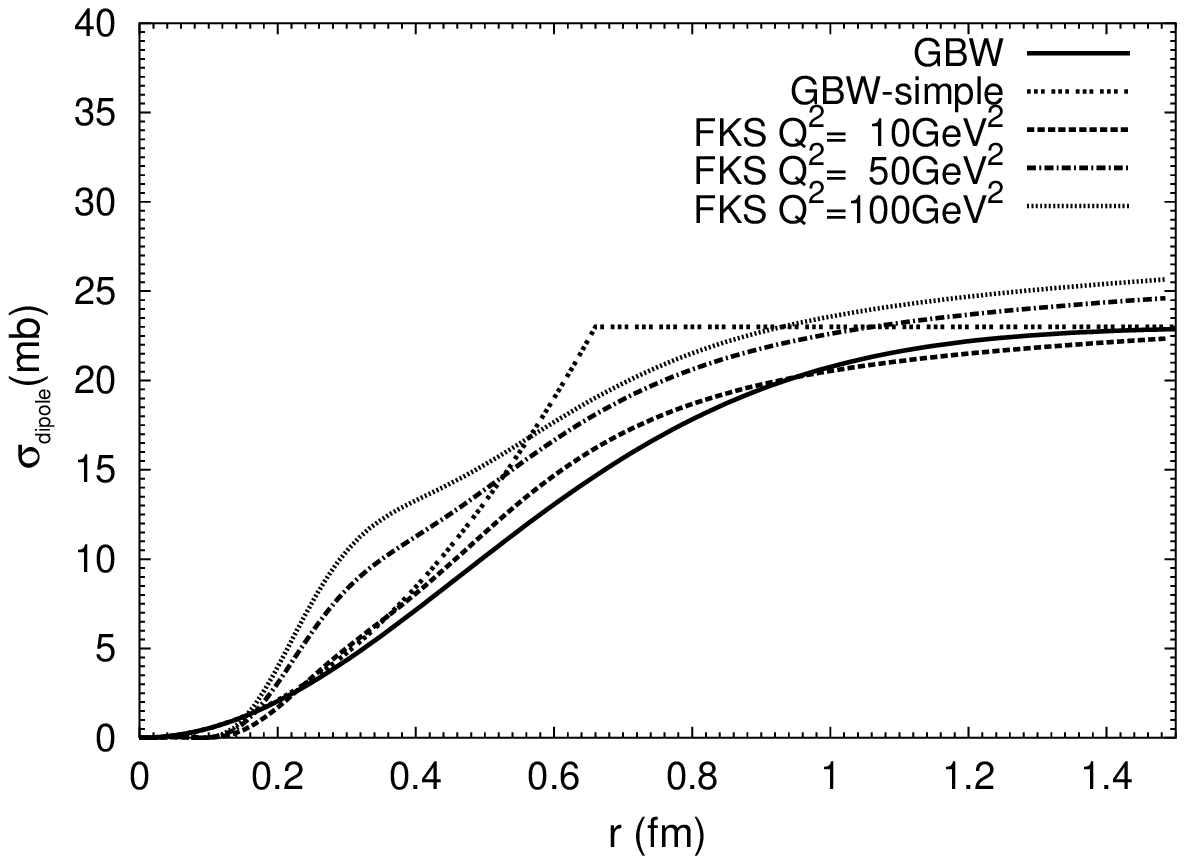,width=12cm,height=10cm}}
\vskip 0.5cm {\caption{Different dipole-proton cross sections as
a function of the dipole size $r$. The full drawn curve is the
Golec-Biernat-Wuesthoff cross section \protect\cite {Golec1} at
$x_0=10^{-2}$,the dashed curves are the Forshaw et al. 
\protect\cite
{Forshaw} cross sections at the same $x_0=10^{-2}$ but different
$Q^2=10 $GeV$^2,50 $GeV$^2$ and $100$ GeV$^2$. The curve marked
(GBW-simple) gives the approximation in Eq.~(\ref{sgbw}) to the
GBW cross section. }}
\end{figure}


\begin{references}

\bibitem{Martin:2002dr}
A.~D.~Martin, R.~G.~Roberts, W.~J.~Stirling and R.~S.~Thorne,
arXiv:hep-ph/0201127.

\bibitem{BFKL}
I.~I.~Balitsky and L.~N.~Lipatov,
Sov.\ J.\ Nucl.\ Phys.\  {\bf 28}, 822 (1978)
[Yad.\ Fiz.\  {\bf 28}, 1597 (1978)];
E.~A.~Kuraev, L.~N.~Lipatov and V.~S.~Fadin,
Sov.\ Phys.\ JETP {\bf 45}, 199 (1977)
[Zh.\ Eksp.\ Teor.\ Fiz.\  {\bf 72}, 377 (1977)].

\bibitem{Bal}
I.~Balitsky,
Nucl.\ Phys.\ B {\bf 463}, 99 (1996)
[arXiv:hep-ph/9509348].

\bibitem{Alt}
V.~S.~Fadin and L.~N.~Lipatov,
Phys.\ Lett.\ B {\bf 429}, 127 (1998)
[arXiv:hep-ph/9802290].
M.~Ciafaloni and G.~Camici,
Phys.\ Lett.\ B {\bf 430}, 349 (1998)
[arXiv:hep-ph/9803389].

\bibitem{ABF}
G.~Altarelli, R.~D.~Ball and S.~Forte,
arXiv:hep-ph/0104246.

\bibitem{Mueller}
L.~V.~Gribov, E.~M.~Levin and M.~G.~Ryskin,
Phys.\ Rept.\  {\bf 100}, 1 (1983);
Y.~V.~Kovchegov and A.~H.~Mueller,
Phys.\ Lett.\ B {\bf 439}, 428 (1998) [arXiv:hep-ph/9805208].


\bibitem{Bartels}
J.~Bartels and C.~Ewerz,
JHEP {\bf 9909}, 026 (1999) [arXiv:hep-ph/9908454].


\bibitem{PirnerPLB}H.~J.~Pirner,
Phys.\ Lett.\ B {\bf 521}, 279 (2001).


\bibitem{NPFV}
H.~W.~Naus, H.~J.~Pirner, T.~J.~Fields and J.~P.~Vary,
Phys.\ Rev.\ D {\bf 56}, 8062 (1997) [arXiv:hep-th/9704135].



\bibitem{HJP}
E.~M.~Ilgenfritz, Y.~P.~Ivanov and H.~J.~Pirner,
Phys.\ Rev.\ D {\bf 62}, 054006 (2000) [arXiv:hep-ph/0003005].

\bibitem{Wilczek}
D.~J.~Gross and F.~Wilczek,
Phys.\ Rev.\ D {\bf 9}, 980 (1974).

\bibitem{Politzer}
H.~D.~Politzer,
Phys.\ Rept.\  {\bf 14}, 129 (1974).

\bibitem{Prokhvatilov}
E.~V.~Prokhvatilov and V.~A.~Franke,
Sov.\ J.\ Nucl.\ Phys.\  {\bf 49}, 688 (1989)
[Yad.\ Fiz.\  {\bf 49}, 1109 (1989)].



\bibitem{Lenz2}
F.~Lenz, H.~W.~Naus and M.~Thies,
Annals Phys.\  {\bf 233}, 317 (1994).


\bibitem{I}
E.~Iancu, A.~Leonidov and L.~McLerran,
arXiv:hep-ph/0202270.

\bibitem{Lenz1}
F.~Lenz, M.~A.~Shifman and M.~Thies,
Phys.\ Rev.\ D {\bf 51}, 7060 (1995)
[arXiv:hep-th/9412113].


\bibitem{beg} M.A.B.~Meg and H. Ruegg, J. Math. Phys. {\bf 6}, 677
  (1965).

\bibitem{Burkardt1}
M.~Burkardt and Y.~Koike,
arXiv:hep-ph/0111343.

\bibitem{Brodsky}
S.~J.~Brodsky, H.~C.~Pauli and S.~S.~Pinsky,
Phys.\ Rept.\  {\bf 301}, 299 (1998) [arXiv:hep-ph/9705477].

\bibitem{Burkardt2}
M.~Burkardt,
Adv.\ Nucl.\ Phys.\  {\bf 23}, 1 (1996)
[arXiv:hep-ph/9505259].

\bibitem{Svetitsky}
B.~Svetitsky and L.~G.~Yaffe,
Nucl.\ Phys.\ B {\bf 210}, 423 (1982).

\bibitem{Karsch1}
F.~Karsch and S.~Stickan,
Phys.\ Lett.\ B {\bf 488}, 319 (2000)
[arXiv:hep-lat/0007019].

\bibitem{Karsch2}
F.~Karsch, C.~Schmidt and S.~Stickan,
arXiv:hep-lat/0111059.

\bibitem{pirner2}
Hans J. Pirner and Feng Yuan,
Nucl.\ Phys.\ Proc.\ Suppl.\ {\bf 108}, 313 (2002).

\bibitem{engel} J.~Engels, private communication.

\bibitem{Golec1}
K.~Golec-Biernat and M.~Wusthoff,
Phys.\ Rev.\ D {\bf 59}, 014017 (1999) [arXiv:hep-ph/9807513];
Phys.\ Rev.\ D {\bf 60}, 114023 (1999) [arXiv:hep-ph/9903358].

\bibitem{H1}
I.~Abt {\it et al.}  [H1 Collaboration],
Nucl.\ Phys.\ B {\bf 407}, 515 (1993);
T.~Ahmed {\it et al.}  [H1 Collaboration],
Nucl.\ Phys.\ B {\bf 439}, 471 (1995) [arXiv:hep-ex/9503001];
S.~Aid {\it et al.}  [H1 Collaboration],
Nucl.\ Phys.\ B {\bf 470}, 3 (1996) [arXiv:hep-ex/9603004];
C.~Adloff {\it et al.}  [H1 Collaboration],
Nucl.\ Phys.\ B {\bf 497}, 3 (1997) [arXiv:hep-ex/9703012].

\bibitem{ZEUS}
M.~Derrick {\it et al.}  [ZEUS Collaboration],
Z.\ Phys.\ C {\bf 72}, 399 (1996) [arXiv:hep-ex/9607002];
Phys.\ Lett.\ B {\bf 316}, 412 (1993);
Z.\ Phys.\ C {\bf 65}, 379 (1995);
Z.\ Phys.\ C {\bf 69}, 607 (1996) [arXiv:hep-ex/9510009];
J.~Breitweg {\it et al.}  [ZEUS Collaboration],
Phys.\ Lett.\ B {\bf 407}, 432 (1997) [arXiv:hep-ex/9707025].

\bibitem{Golec2}
K.~Golec-Biernat,
arXiv:hep-ph/0109010.


\bibitem{Golec3}
A.~M.~Stasto, K.~Golec-Biernat and J.~Kwiecinski,
Phys.\ Rev.\ Lett.\  {\bf 86}, 596 (2001)
[arXiv:hep-ph/0007192].


\bibitem{Forshaw}
J.~R.~Forshaw, G.~Kerley and G.~Shaw,
Phys.\ Rev.\ D {\bf 60}, 074012 (1999)
[arXiv:hep-ph/9903341].

\bibitem{Lipatov:1990zb}
L.~N.~Lipatov,
Phys.\ Lett.\ B {\bf 251} (1990) 284
[Nucl.\ Phys.\ Proc.\ Suppl.\  {\bf 18C} (1990) 6].


\bibitem{Ewerz:2001uq}
C.~Ewerz,
Phys.\ Lett.\ B {\bf 512} (2001) 239
[arXiv:hep-ph/0105181].

\end{references}
\end{document}